%% file: main.tex
\documentclass[fleqn,usenatbib]{mnras}

\usepackage{newtxtext,newtxmath}
\usepackage[T1]{fontenc}
\usepackage{tabularx}
\usepackage[normalem]{ulem}
\usepackage{multirow}
\usepackage{makecell}
\usepackage{textcomp}
\usepackage{subcaption}
\usepackage{dcolumn}
\usepackage{multirow}
\usepackage{color}
\usepackage{hyperref}
\usepackage{url}

\DeclareRobustCommand{\VAN}[3]{#2}
\let\VANthebibliography\thebibliography
\def\thebibliography{\DeclareRobustCommand{\VAN}[3]{##3}\VANthebibliography}

\usepackage{graphicx}	
\usepackage{amsmath}	
\hypersetup{
    colorlinks=true,
    linkcolor=blue,
    citecolor=blue,
    urlcolor=blue,
} 

\title[Supernova Gravitational Waves with Machine Learning]{Exploring Supernova Gravitational Waves with Machine Learning}
\input{formula.tex}

\author[A. Mitra et al.]{
A. Mitra,$^{1,2,3}$\thanks{E-mail: ayan.mitra@nu.edu.kz}
B. Shukirgaliyev,$^{4,5,6}$
Y. S. Abylkairov,$^{4,7}$
and E. Abdikamalov$^{1,4}$
\\
$^{1}$Department of Physics, Nazarbayev University, 53 Kabanbay Batyr ave, 010000 Astana, Kazakhstan\\
$^{2}$The Inter-University Centre for Astronomy and Astrophysics (IUCAA), Post Bag 4, Ganeshkhind, Pune 411007, India\\
$^{3}$Kazakh-British Technical University, Almaty, Kazakhstan\\
$^{4}$Energetic Cosmos Laboratory, Nazarbayev University, 53 Kabanbay Batyr ave, 010000 Astana, Kazakhstan\\
$^{5}$Fesenkov Astrophysical Institute, 23 Observatory str, 050020 Almaty, Kazakhstan\\
$^{6}$Faculty of Physics and Technology, Al-Farabi Kazakh National University, 71 Al-Farabi ave, 050020 Almaty, Kazakhstan\\
$^{7}$Department of Mathematics, Nazarbayev University, 53 Kabanbay Batyr ave, 010000 Astana, Kazakhstan
}

\date{Accepted XXX. Received YYY; in original form ZZZ}

\pubyear{2022}

\begin{document}
\label{firstpage}
\pagerange{\pageref{firstpage}--\pageref{lastpage}}
\maketitle

\begin{abstract}
Core-collapse supernovae (CCSNe) emit powerful gravitational waves (GWs). Since GWs emitted by a source contain information about the source, observing GWs from CCSNe may allow us to learn more about CCSNs. We study if it is possible to infer the iron core mass from the bounce and early ring-down GW signal. We generate GW signals for a range of stellar models using numerical simulations and apply machine learning to train and classify the signals. We consider an idealized favorable scenario. First, we use rapidly rotating models, which produce stronger GWs than slowly rotating models. Second, we limit ourselves to models with four different masses, which simplifies the selection process. We show that the classification accuracy does not exceed $\sim\! 70\%$, signifying that even in this optimistic scenario, the information contained in the bounce and early ring-down GW signal is not sufficient to precisely probe the iron core mass. This suggests that it may be necessary to incorporate additional information such as the GWs from later post-bounce evolution and neutrino observations to accurately measure the iron core mass. 
\end{abstract}

\begin{keywords}
Gravitational Waves -- Supernovae: general
\end{keywords}


\section{Introduction}

The detection of gravitational waves (GWs) from  black hole mergers gave birth to    GW astronomy \citep{abbot16PRL}, while the observation of neutron star merger in gravitational and electromagnetic waves is the prime  example of the multi-messenger astronomy \citep{abbott17PRL}. One of the most promising multi-messenger sources that are yet to be detected in GWs are core-collapse supernovae (CCSNe). 

CCSNe are the powerful explosions of massive stars at the end of their life. The collapse of the stellar core releases $\sim\! 10^{53}$ erg of its gravitational binding energy. While most of this escapes in the form of neutrinos, the rest powers the explosion \citep{Bethe:1990mw}. Supernovae involve powerful aspherical flows that generate GWs with energies up to $\sim\! 10^{47}$ erg \citep[e.g.,][]{ott:09review, kotake:13review, kotake:17}. The explosion front breaks out of the stellar surface hours later \citep{waxman17}, producing a blast of photons across the electromagnetic spectrum \citep[e.g,][]{Nakamura:2016kkl}.

The supernova dynamics, and thus the GW signal, strongly depends on rotation. In non- or slowly rotating models, which represent the majority of CCSNe \cite[e.g.,][]{heger:05}, the explosion is governed by the \emph{neutrino mechanism} \citep[e.g.,][for recent reviews]{kotake:12snreview, janka:12a, radice:18, mueller:20review, Burrows21review, mezzacappa22}. A fraction of neutrinos emitted by the protoneutron star (PNS) heats the post-shock material, giving rise to neutrino-driven convection \citep{herant:92, bhf:95, janka:95} and standing accretion shock instability (SASI) \citep{blondin:03, foglizzo:06, mueller:12b}. These flows perturb the PNS and excite its oscillations, generating powerful GWs \cite[e.g.,][]{murphy:09, mueller:13,cerda:13, yakunin15, andresen:17, Hayama18, radice:19gw, mezzacappa:20a, Raynaud22}. Asymmetries in neutrino emission also contribute to the GW signal \cite[e.g.,][]{Kotake07, takiwaki:18, Vartanyan20}. The GWs are detectable for sources within our Galaxy with current generation detectors \citep{gossan:16, abbott:20ccsn, Lopez21, Szczepanczyk21, Antelis22}, while the future generation detectors will enable a more detailed observation { or observations of more distant supernovae} \citep{Srivastava:2019fcb, powell:19, powell:20}.

In rare rapidly rotating stars, the rotational kinetic energy powers the explosion via the \emph{magnetorotational mechanism} \citep{leblanc:70, bisno:76, akiyama:03, burrows:07a, moesta:14a, kuroda:20, obergaulinger:20, Raynaud20}. Due to the centrifugal force, the PNS forms with an oblate perturbation, triggering PNS ring-down oscillations that last for $\sim\! 10 $ ms \citep{ott:12, fuller:15}. In some cases, the PNS may be subject to non-axisymmetric instabilities \citep[e.g.][]{ott:07cqg, scheidegger:10cqg, shibagaki:20, Takiwaki21}. When this happens, the non-axisymetrically-deformed PNS emits GW for many rotation periods, significantly enhancing the detectability \citep{abbott:20ccsn}. For moderate rotation, both the rotational bounce and convection contribute to the GW signal \citep{andresen19, Pan21, Jardine22}.

Once detected, it is possible to estimate the parameters of source using the GWs emitted by the source \citep[e.g.,][for a recent review]{Christensen22}. Since slowly and rapidly rotating models have significantly different dynamics, their GW signals can be confidently distinguished \citep{Logue12, Powell16, Chan20, Szczepanczyk21, Saiz-Perez22}. Both convection and SASI develop from stochastic perturbations, so the GWs coming from slowly rotating models contain stochastic components. In contrast, the bounce GW signal in rapidly rotating stars can be determined precisely for a given (physical and computational) model parameters \cite[e.g.,][]{zwerger:97, dimmelmeier:08}. Despite the presence of the stochastic contributions, the GW spectra contain the frequencies of the physical processes happening in the central regions \citep[e.g.,][]{Kotake11, mueller:13, Astone18, Roma19, Srivastava:2019fcb, Powell22}. In particular, the PNS oscillations is the dominant component of the signal \citep{murphy:09, cerda:13, morozova:18, radice:19gw, Warren20}, the frequencies of which can be related to the physical parameters of the system, such as the mass and radius of the PNS \cite[e.g.,][]{dimmelmeier:06, mueller:13, Torres-Forne19, Vartanyan19, pajkos19, Sotani21, Bizouard21}. 

For rapidly rotating stars, \citet{abdikamalov:14} explored the possibility to infer the rotation and its distribution in the supernova core. They considered $\sim\!100$ different rotational configurations of a stellar model with 5 different degrees of differential rotation. They found that, for a source at $10$~kpc distance, it is possible to measure rotation with $\sim\! 20\% $ accuracy for rapidly rotating models, in which the rotational kinetic energy exceed $\sim\! 8\%$ of the potential binding energy. In slowly rotating models, due to smaller GW amplitudes and higher stochastic contribution, the error becomes larger. These estimates were further improved by \citet{Engels14, Edwards14}, and \citet{Afle21}. \citet{Hayama16} showed that signs of rapid rotation can be found in the circular polarization of the GW signal. \citet{Yokozawa15} proposed to combine GW and neutrino observations to infer the rotation from the time delays between the bursts of these two signals. \citet{pajkos19, pajkos21} proposed to combine the the core-bounce signal with the dominant frequency mode of the PNS in the pre-explosion to constrain the structure of the progenitor star. 

\citet{richers:17} studied the dependence of the GW signal on the parameters of the equation of state (EOS) of high-density matter in protoneutron stars and treatment the electron capture rate during collapse. They find a modest impact of these parameters to the bounce and the early $\sim\! 20$ ms post-bounce GW signal. This emphasizes the importance of accurate modeling \citep[e.g.,][]{lentz:12a, Kotake18, Pan18, Pan19, mezzacappa20, Andresen21}. Using deep convolutional neural networks, \citet{Edwards21} classified these EOSs $72\%$ correctly, while their most probable five EOSs were found with $97\%$ accuracy \cite[see also][]{Chao22}. 

In this work, we study if it is possible to probe the iron core mass of CCSN progenitor from the bounce and the early { $\sim 10$ post-bounce oscillations of the newly-formed PNS before it settles in a quasi-equilibrium state. Following the convention used in the literature \cite[e.g.,][]{abdikamalov:22review}, we will refer to these oscillations ring-down oscillations.} \citet{ott:12} showed that progenitors with different masses will produce similar GWs if they have a similar angular momentum distributions at a given mass coordinate in the stellar core. This suggest that the progenitor mass at most has a subtle effect on the bounce dynamics. We extend this work further by using machine learning (ML) for signal classification and studying a wider ranges of progenitors and rotational configurations. We look at idealized optimistic scenario for measuring mass from the bounce and early ring-down GW signal alone. First, we consider rapidly rotating models, which produce strong GWs. Second, we use the deleptonization method that is known to artificially amplify the differences between progenitors. Third, we limit ourselves to model progenitors with four different iron core masses only, which simplifies the selection process. Despite these ideally favorable conditions, we show that the iron core mass cannot be accurately measured from the GW bounce signal alone. This suggests that, to measure the mass, one has to incorporate additional information such as longer post-bounce signal or neutrino observations, or both. 


\section{Methodology}

\subsection{Gravitational waveforms}
\label{sec:gw_data}

We consider four progenitor models with zero-age main sequence masses ranging from $12$ to $40$ $M_\odot$. At the pre-collapse stage, these models develop iron cores with masses ranging from $1.3M_\odot$ to $1.8M_\odot$, as shown in Table ~\ref{tab:progenitor_param}, which also provides the central densities and entropies. The $12M_\odot$ and $40M_\odot$ models are produced by \cite{woosley:07}, while the $15M_\odot$ model is evolved by \cite{heger:05} with magnetic field prescription by \cite{spruit:02}. The $27M_\odot$ model is produced by \cite{whw:02}. All models have solar metallicity. 

To obtain the GW signals, we perform simulations using the {  general relativistic hydrodynamics code} {\tt CoCoNuT} \citep{Dimmelmeier02a, Dimmelmeier02b, dimmelmeier:05MdM}. { {\tt CoCoNuT} employs the conformal-flatness condition (CFC), which is an excellent approximation to full general relativity for modeling stellar core collapse \cite[e.g.,][]{ott:07cqg}. We perform simulations until 25 ms after bounce. This time interval contains the bounce and ring-down oscillations of the PNS, which emit most of the GW signal in this phase \citep[][]{abdikamalov:22review}, while remaining largely axisymetric \citep{ott:07cqg}. For this reason, we perform our simulations in 2D axisymmetry. Our computational domain spans a radius of $3000$ km covered with 250 logarithmically spaced radial cells with the central resolution of $250$ m. We assume equatorial symmetry. The upper $90^\circ$ of our domain is covered with 40 angular cells of uniform size. We have performed extensive resolution tests to verify that the adopted resolution is sufficient for modeling the GW signal in this phase \citep[e.g.,][]{abdikamalov:14}.}

Following \citet{richers:17}, we use the SFHo nuclear equation of state \citep{steiner:13b}. During collapse, our code uses the $Y_\mrm{e}(\rho)$ deleptonization method of \cite{liebendoerfer:05}. Following \citet{richers:17}, we use $Y_\mrm{e}(\rho)$ profile obtained from spherically symmetric radiation hydrodynamics simulations using the {\tt GR1D} code \citep{oconnor:15a}. This method is known to amplify the contrast between progenitors with different masses \citep{mueller:09phd, pajkos21}: it produces variations in the inner core mass at bounce of $\simeq 10\%$ between progenitors with different masses, whereas full neutrino-transport simulations yields inner core masses { at bounce} that are practically independent of the progenitor \citep{mueller:09phd, pajkos21}\footnote{ While simulations with advanced neutrino transport show that the mass of the inner core at bounce is independent of the progenitor \citep{mueller:09phd, pajkos21}, this is not the case for the PNS mass in the post-bounce phase, which does depend on the progenitor properties \cite[e.g.,][]{bruenn:16, mueller:16c, ott:18, burrows19, nagakura:20}}. However, this weakness of the method represents the strength of our work: as we show below, even with the artificial amplification, the difference in the GW signals is too small to be distinguishable. In more realistic models, the distinction should be even smaller{ , which makes it even harder to distinguish the mass}.  

Following \cite{abdikamalov:14}, for each progenitor we consider about $100$ different rotational profiles ranging from slow to rapid rotation, where rotation has little to large effect on the core dynamics, respectively. At distance $\varpi$ from the rotation axis, the angular velocity of pre-collapse models is given by
\begin{equation}
    \label{eq:rot_law}
    \Omega(\varpi) = \Omega_0 \left[ 1 + \left(\frac{\varpi}{A}\right)^2\right]^{-1},
\end{equation}
where $A$ is a measure of degree of differential rotation, $\Omega_0$ is the central angular velocity \citep{komatsu:89a}. As in \cite{abdikamalov:14}, we consider five different values of $A$, ranging from $300$ to $10,\!000$ km, which correspond to the limits of, respectively, extreme differential and uniform rotations in the stellar core. For a given $\Omega_0$ and $A$, we impose the rotation law (\ref{eq:rot_law}) to the $12M_\odot$ model, as described in \cite{abdikamalov:14}. We then map the specific angular momentum distribution homologously to the other progenitors with different masses. This ensures that the specific angular momentum at a given enclosed mass coordinate is similar in all four progenitors. As summarized in Table~\ref{tab:T1}, the total number of rotational configurations we consider is 97, 99, 102, and 104 for s12, s15, s27, and s40 models, respectively. The difference in this number is caused by the fact that at the extreme rapid rotation limit, due to a fine balance between gravity and centrifugal force in the pre-collapse stage, some of the models do not collapse for some progenitors and collapse for other. 

We extract the GW waveforms from the simulations using the Newtonian quadrupole formula in the first moment of momentum density formulation \citep{Dimmelmeier02a, Dimmelmeier02b, dimmelmeier:05MdM}, which yields accurate waveforms for stellar core collapse simulations \citep{reisswig:11ccwave}.

\begin{table}
        \setlength\extrarowheight{2.09pt}
        \begin{center}
            \begin{tabular}{|c|c|c|c|}
                \hline
                ZAMS mass & $M_\mrm{IC}$ & $s_\mrm{c}$ & $\rho_\mrm{c}$  \\ [.01em]
                [$M_\odot$] & [$M_\odot$] & [$k_\mrm{B}/\mrm{baryon}$] & [$10^9\,\mrm{g/cm}^3$] \\ [.3em]
                \hline
                12   & 1.3 & $0.60$ & $12.0$ \\ 
                15   & 1.4 & $0.62$ & $8.74$ \\ 
                27   & 1.5 & $0.77$ & $4.96$ \\
                40   & 1.8 & $1.06$ & $2.17$ \\
                \hline
            \end{tabular}
        \end{center}
        \caption{Progenitor model parameters used in our work. The ZAMS mass is the initial (the so-called zero-age) mass of the star. Parameter $M_\mrm{IC}$ is the inner core mass, while $s_\mrm{c}$ and $\rho_\mrm{c}$ are the central specific entropy and density before collapse.}
        \label{tab:progenitor_param}
    \end{table}

\begin{table}
        \setlength\extrarowheight{2.09pt}
        \begin{center}
            \begin{tabular}{c c}
                \hline
                Progenitor Mass & No. of Events \\ [.3em]
                [$M_\odot$]     &  \\ [.3em]
                \hline
                12 & 97 \\ 
                15 & 99 \\ 
                27 & 102 \\
                40 & 104 \\
                \hline
               Total & \total\\
                \hline
            \end{tabular}
        \end{center}
        \caption{Number of the generated GW events grouped by their progenitor mass types.}
        \label{tab:T1}
    \end{table}


\subsection{Machine learning}

\begin{figure*}
    \centering
    \makebox[1.\textwidth]{\includegraphics[width=1.0\textwidth]{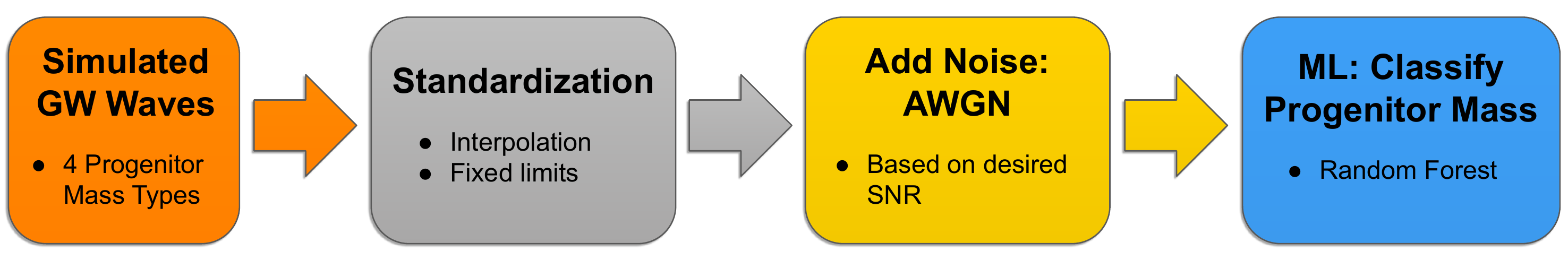}}
    \caption{Flowchart of the analysis pipeline of our study.}
    \label{fig:flowchart}
\end{figure*}

Our analysis pipeline can be divided in the following steps:
\begin{itemize}
    \item Preprocessing: Customize the waveforms for ML input.
    \item Noise: Add noise to the signals to model the detector noise.  
    \item ML: Optimize hyperparameters\footnote{Hyperparameters are  parameters of the algorithm, which are tuned to optimize the learning. Their values are chosen before initialising learning. They should not be confused with parameters of the data, which are obtained during the learning. The choice of hyperparameter has a direct impact on the performance of the ML model and hence they are optimized to produce the best performance.} and apply ML algorithm.
    \item Analysis: Explore confusion matrix and other metrics. 
\end{itemize}
A flowchart depiction of the pipeline is illustrated in Fig.~\ref{fig:flowchart}. 

To make the waveforms suitable to ML analysis, we condense all candidate waveforms into a single data cube with the progenitor mass of each candidate treated as target class. We map each candidates to the same cadence along the time axis with uniform starting and ending points of $-5$ and $+15$ ms around the bounce time. The time of bounce corresponds to time equal to zero. We generate $10,000$ interpolated points along the time axis per candidate, resulting in a data cube of dimensions $(\total\times10000\times2)$. 


\subsubsection{Random Forest}
\label{sec:rf}

\begin{table}
        \setlength\extrarowheight{2.09pt}
        \begin{center}
            \begin{tabular}{|c | c| }
                \hline
                hyperparameters & Range   \\ [.3em]
                \hline \hline
                No. of estimators & $[100,400]$   \\ 
                Max. features & [auto, sqrt, log2] \\ 
                Max. depth  & $[10,250]$\\
                Min. samples split & $[2,100]$ \\
                min. samples leaf & $[2,100]$\\
                Bootstrap  & [True, False]\\
                OOB score  & [True, False]\\
                Warm start & [True, False] \\
                \hline
                \hline
            \end{tabular}
        \end{center}
        \caption{Hyperparameters and their ranges (or states) used during random grid search.}
        \label{tab:rf_hyper}
    \end{table}

\begin{figure}
    \includegraphics[width=\columnwidth]{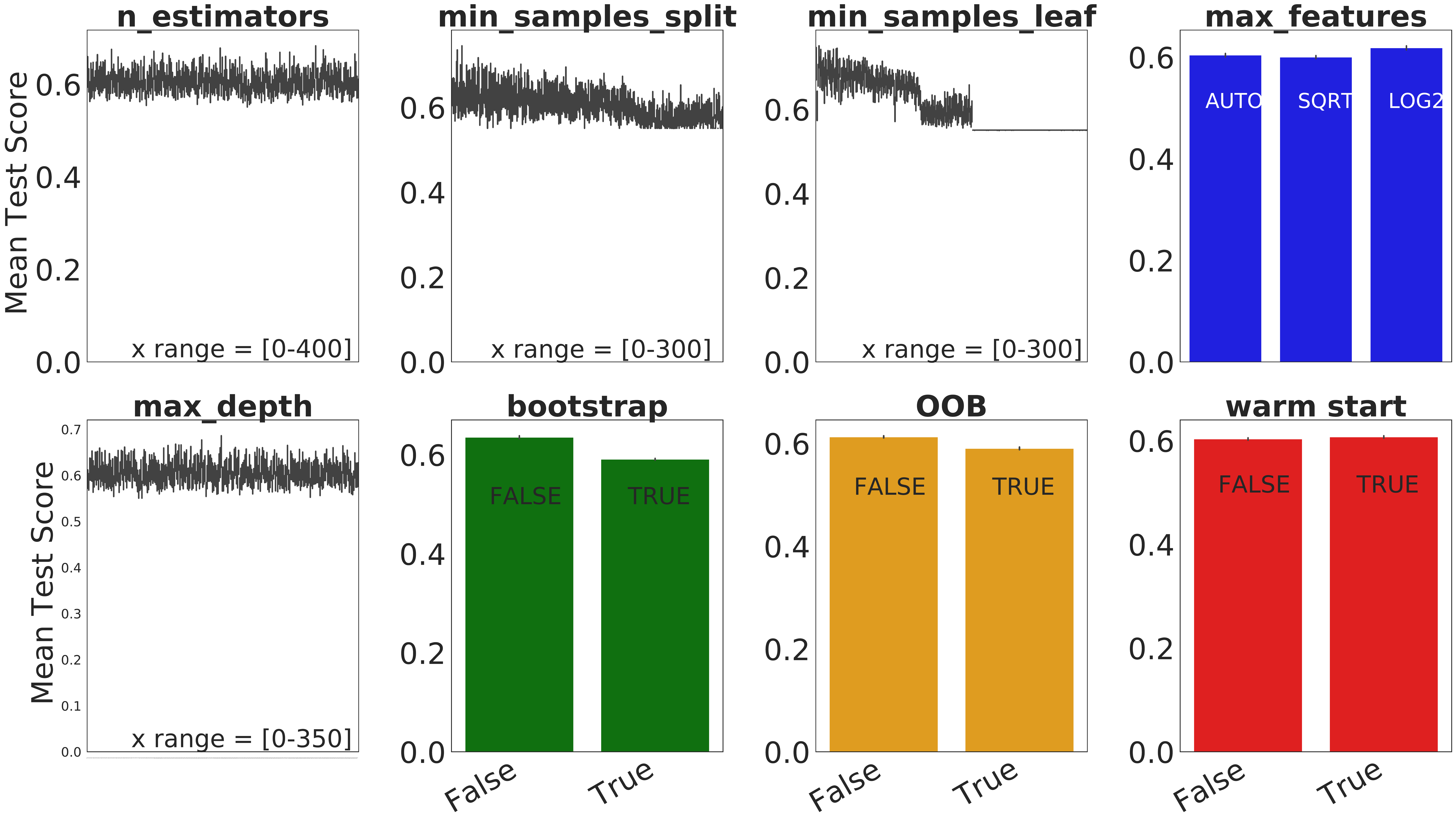}
    \caption{Mean test score of grid search results for $\mathrm{SNR}=100$ plotted against individual hyperparameters listed in Tab.~\ref{tab:rf_hyper}.}
    \label{fig:hyper}
\end{figure} 

Random forest (RF) is a learning method that operates by constructing an ensemble of decision trees \citep{tree}. It can be applied to classification and regression problems. For classification problems, the RF output is the class selected by the largest number trees, while for regression problems, the result is the average prediction of all trees. RF classifies samples using a forking path of decision points. We move from one decision point to another and at each point we apply a rule that decides which branch to follow \citep{Rf-advntg1}. At the end, we arrive at a leaf, which has a corresponding class label. We conclude the selection process by assigning the data to the class \citep{Denisko18}.

Tree-based approaches such as the RF have generated significant interest recently \citep{kennedy,hernandez,bonjean, Tsang22} because they are robust, difficult to overfit, requires minimal computational resources, and has availability of methods to assist in interpreting the results \citep{RF-Hastie2010,RF-breiman2001}, while being quick to implement. It is also efficient in learning highly non-linear relations between the input and the labels, making them suitable for large training data set and large number of input variables.

We use the accuracy score as the performance metric for optimizing our RF classifier network and to evaluate their performance. To arrive at the final classifier settings for running the pipeline, we split our train-test data into $90-10\%$ ratio and tune several hyperparameters using a randomised grid search algorithm \citep{sklearn_api}. Table~\ref{tab:rf_hyper} lists all the hyperparameters used in the pipeline and their values (or ranges) considered during the grid search. We perform a $100$ fold randomized cross validation for tuning the hyperparameters. Since cross validation can produce an unbiased estimator of a model on unseen data \citep{kuhn2013applied}, we do not use a separate validation dataset. For comparison and consistency of our results, we test additional algorithms, as discussed in Appendix \ref{appendix-1}).


\subsubsection{Detector Noise}
\label{sec:noise}

The GW detectors can be impacted by a wide variety of noises, stemming from, e.g., quantum sensing, seismic disturbances, suspension and mirror coating thermal vibrations, and transient perturbations such as anthropogenic interventions, weather, equipment malfunctions as well as occasional transient noise of unknown origin \citep{LIGOplus, noise-1, noise-2, Driggers19, noise-3, Ormiston20}. In this analysis, we model the combination of all these possible sources of noise as additive white Gaussian noise (AWGN) \citep{awgn1,awgn0,awgn2}. While AWGN cannot model instrumental noise transients and loud GW bursts contributing to non-Gaussian and non-stationary features \citep{Abbot20ligonoise}, as we show below, this approximation is adequate to establish the upper limit for the classification accuracy, which is the main goal of this work. 

For a discrete time sequence signal, $x(t_i) \equiv x_i$, the continuum signal $G(t)$ can be defined with the help of a linear correlator as 
\begin{equation}
    G \equiv \int q(t)x(t)dt,
    \label{eq:noise}
\end{equation}
where $q(t)$ encapsulates the information from the inverse noise covariance matrix. This signal can be matched to a known waveform from a linear matched filter. For zero signal, the expectation value $E$ of $G(t)$ becomes zero. One can define the signal to noise ratio (SNR) as 
\begin{equation}
    \frac{S}{N}=\frac{ E\{G\} }{\sqrt{{\left(\sigma\left\{G\right\}\right)}^2}},
    \label{eq:noise2}
\end{equation}
where the numerator is the expectation\footnote{It is derived from the power spectral density of the signal noise \citep{Flanagan:98}.} and denominator is the variance of the signal. In other words, it can be defined as the ratio of signal power to variance. The SNR is a measure of the strength of a signal observed by a detector with a given level of noise. It is directly proportional to the amplitude of the signal buried in the noise.    

In our  analysis, we add AWGN noise $h_\mathrm{noise}$ to GW strain $h$ in the time domain before the application of ML analysis, for every single chosen value of SNR. The mean value of noise is assumed to be 0. The standard deviation $\sigma_\mathrm{noise}$ of $h_\mathrm{noise}$ is governed by the value of SNR. A consistency check on the stationarity and Gaussianity of the waveform distribution can be done by transforming them in Fourier space and studying their real and imaginary parts individually \citep{Abbot20ligonoise}. If noise distributions for both these components yield mean value of $0$, then white Gaussianity can be assumed. In this analysis, we have performed this check on our noise data and found them to be consistent with this requirement.

As the final step before applying ML, we standardize the dataset using the Z-score normalization \citep{zscore,zscore2}. Roughly speaking, this normalization moves the mean of the signal to $0$ and adjusts the average strength to $1$. 

\section{Results}

\begin{figure}
    \includegraphics[width=\columnwidth]{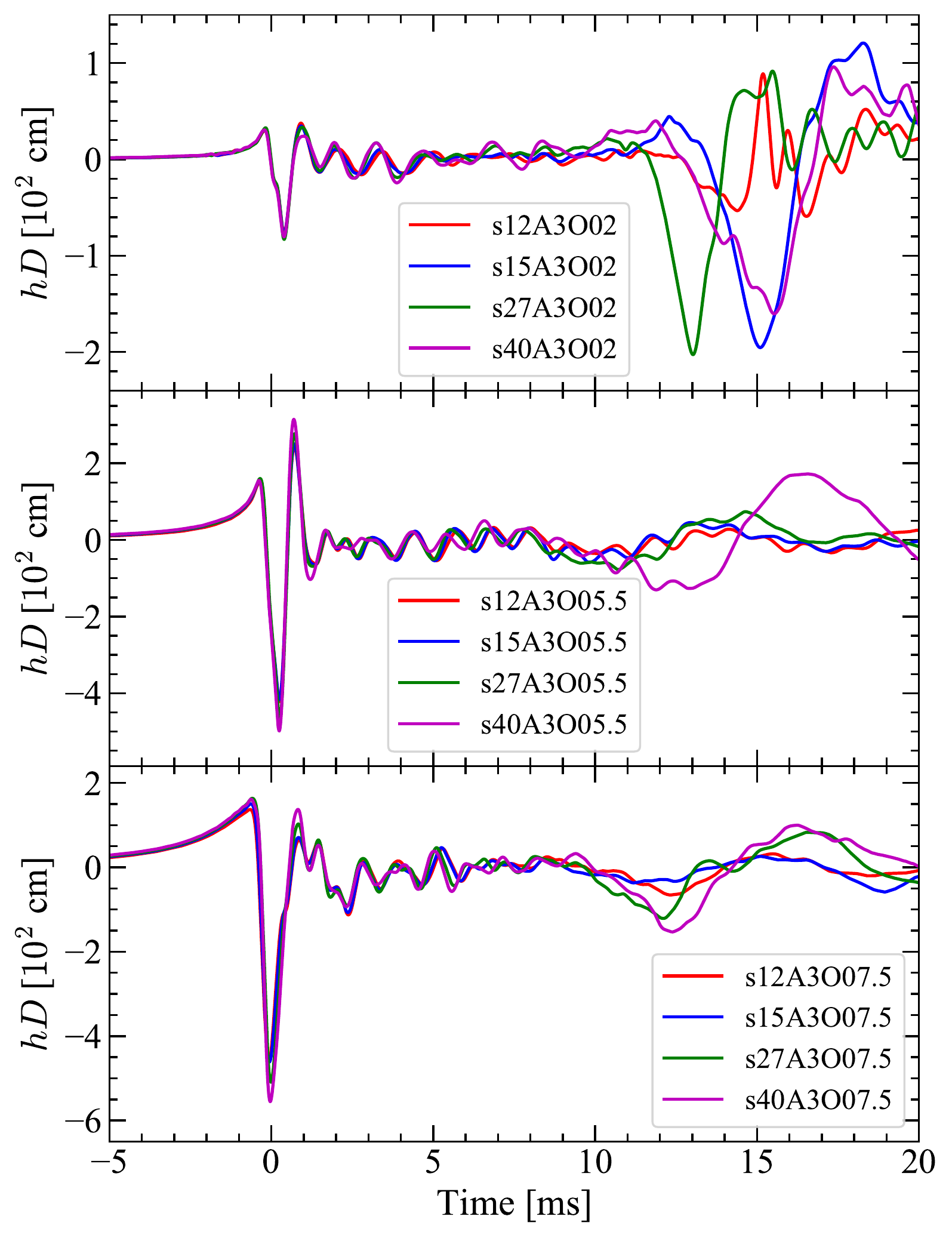}
    \caption{GW strain as a function of time for different progenitors. The top, center, and bottom panels show slowly, moderately rapidly, and rapidly rotating models, respectively. The time value of $0$ ms corresponds to the time of bounce.}
    \label{fig:GW_strain}
\end{figure}

The top, center, and bottom panels of Fig.~\ref{fig:GW_strain} show the GW strain as a function of time for three sets of models with slow, moderate and rapid rotations, respectively. The models in each set have similar distributions of angular momentum in their cores. The plot shows that they produce GW signals that are similar to each other. This supports the observation that different progenitors produce similar bounce GW signal for the same distribution of rotation across mass coordinate \citep{ott:12}. Slowly rotating models exhibit contribution from prompt convection $\gtrsim 6 \, \mathrm{ms}$ after bounce. Since convection is a stochastic process, there is no obvious correlation with the progenitor mass within the time scale considered in this work.

The minor difference in the GW signal between different progenitor masses is caused by the differences in the specific entropy of the iron core. The latter tend to increase with the progenitor mass, but the dependence can be non-monotonic \citep[e.g.,][]{whw:02}. The variation of the inner core mass is amplified by our treatment of deleptonization \citep{mueller:09phd, pajkos21}. As discussed above, our deleptonization scheme produces inner core masses bounce that are $\simeq 10\%$ larger in progenitors with large masses than those in with small masses. In full neutrino transport simulations, the inner core mass is practically independent of the progenitor mass \citep{mueller:09phd, pajkos21}. For this reason, we can only put the upper limit to the accuracy of the classification accuracy, which, as we show below, is too low for any reliable identification of the progenitor mass from GW bounce signal along.

\begin{figure}
    \includegraphics[width=\columnwidth]{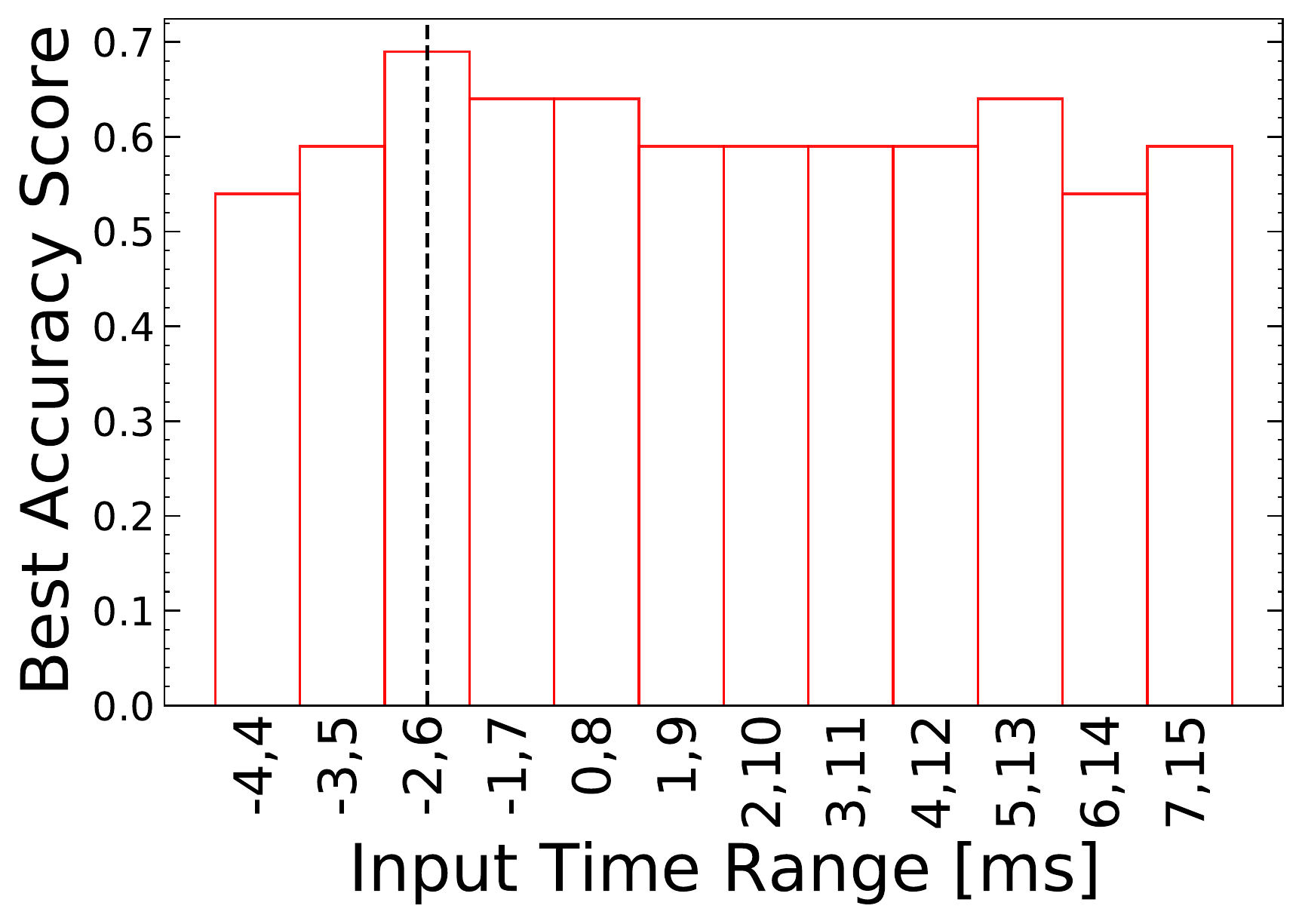}
    \caption{The best accuracy score returned by the RF classifier pipeline for classifying the progenitor mass for different input time ranges using the rolling window of fixed width of $8$ ms. The time value of $0$ ms corresponds to the time of bounce.}
    \label{fig:rolling}
\end{figure} 

Figure~\ref{fig:rolling} shows the best accuracy score for classifying the progenitor mass for different ranges of time windows computed at maximum SNR of $100$. This score measures how many labels the model gets right out of the total number of predictions. The highest accuracy is achieved for range \window, which is also the range used in the analysis of \citet{richers:17}. This is expected since most of the bounce and ring-down GW signal is emitted in this time interval \citep[e.g.,][]{abdikamalov:22review}. Moreover, the prompt convection develops $\gtrsim\! 6$ ms after bounce, which, due to its stochastic nature, makes it harder to identify the source parameters from the signal. Consequently, we carried out the rest of the analysis with the wave spectra in the range \window.

\begin{figure}
    \includegraphics[width=\columnwidth]{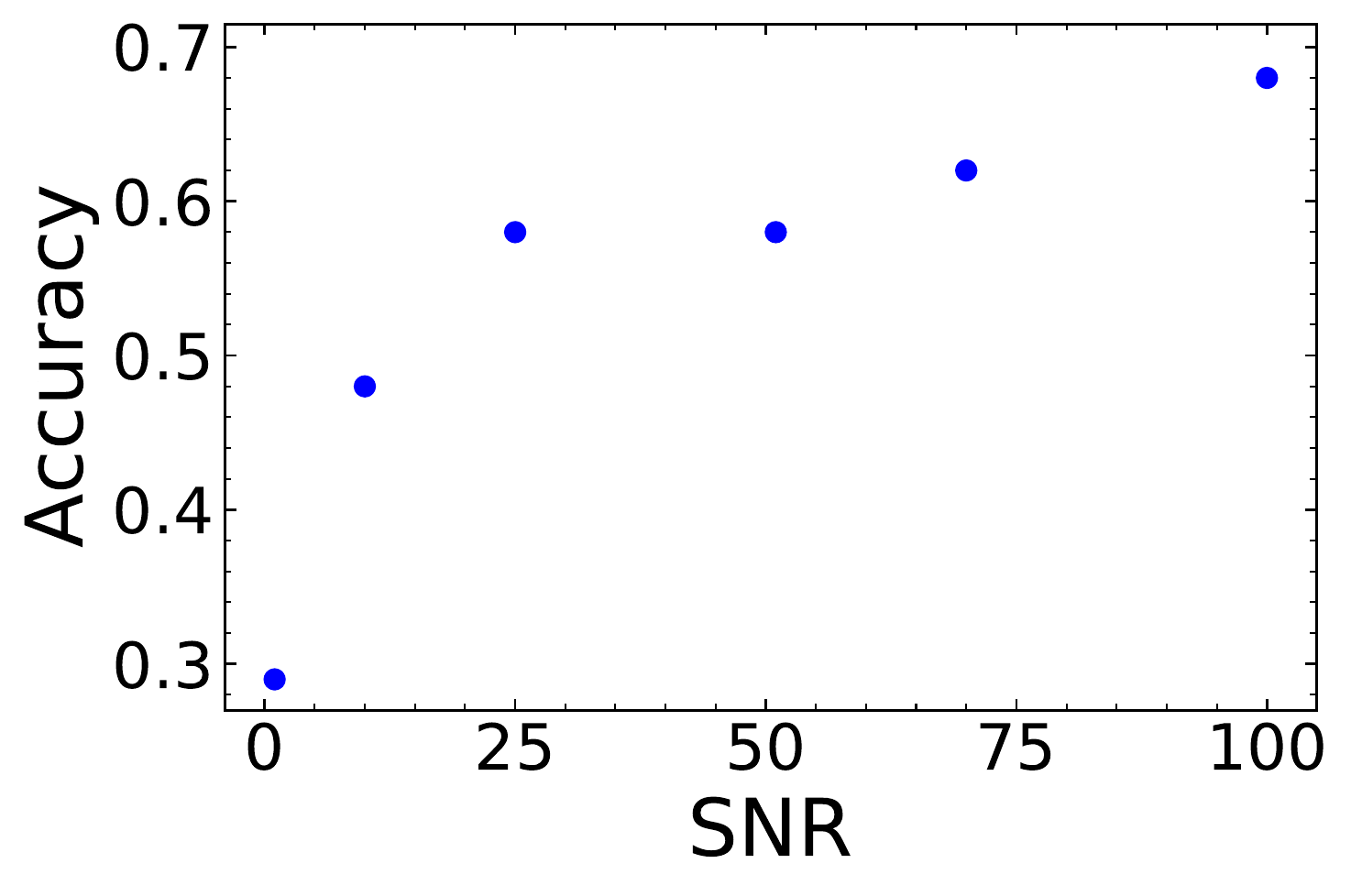}
    \caption{Classifier accuracy score as a function of the signal SNR. It is seen that for low SNR values ($\le30$), accuracy rises sharply with SNR, beyond which accuracy increases marginally with maximum $\sim0.70$ at $\mathrm{SNR}=100$.}
    \label{fig:accuracy}
\end{figure}

We run the above pipeline for six different SNR values $1, 10, 25,  50, 70$, and $100$. We then compute the best accuracy score for each run. This score measures how many labels the model gets right out of the total number of predictions. Fig.~\ref{fig:accuracy} shows the accuracy rate of the classifier as a function of the SNR. The accuracy curve is obtained using cubic interpolation between the discrete SNR values. At the highest SNR of $100$, we see that there is $\sim70\%$ probability of correct classification of the progenitor mass.

\begin{figure}
    \includegraphics[width=\columnwidth]{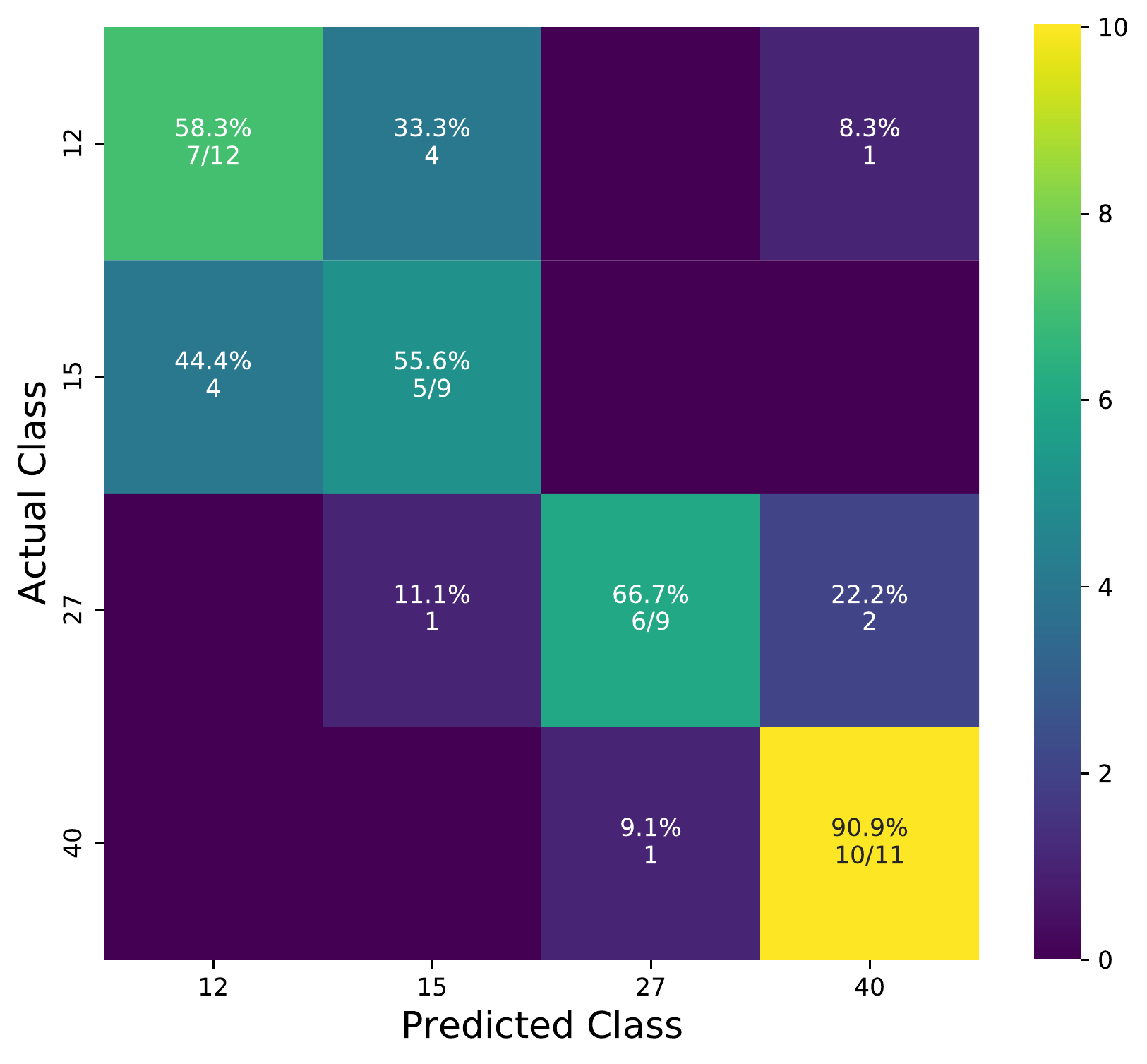}
    \caption{Confusion matrix for $\mathrm{SNR}=100$ presented for test data of $41$ candidates which is $10\%$ of the full strength catalogue used. The fraction of correct prediction per class can be seen along the diagonal.}
    \label{fig:Confusion}
\end{figure}

This finding is mirrored in the behavior of the confusion matrix used to asses the classifier performance shown in Fig.~\ref{fig:Confusion}. Along each rows, the fraction of correct prediction is shown in percentage. The sum of these fractions is $1$. The correct predictions are along the diagonal. All classes have more than $50\%$ correct predictions. The s40 model is the most successfully predicted class with $90.9\%$ cases accurately predicted. { The s15 model has the lowest prediction accuracy of $55.6\%$, while s12 and s27 have the intermediate prediction accuracies of $58.3\%$ and $66.7\%$, respectively.}

{ The explanation of this behavior is simple. In our simulations, the main cause of the differences in the GWs between different progenitors with the same angular momentum distribution is the specific entropy in the iron core. This affects the mass of the inner core at bounce \citep{goldreich:80,yahil:83, burrows:83}, which is a key parameter that determines the properties of the GW signal emitted at bounce and in the ring-down phase \citep[e.g.,][]{dimmelmeier:08}. Since the s40 model has the highest specific entropy in the core prior to collapse (cf. Table~\ref{tab:progenitor_param}), it is relatively easy to distinguish it from the rest of the models. The inner core entropy of the s15 models is close that of the s12 model, which makes it hard to distinguish from that model. Finally, the s27 model has core entropy that is intermediate between the s15 and s40 models, resulting in moderate prediction accuracy. At the same time, as demonstrated by \cite{mueller:09phd}, the dependence on entropy is the artifact of the deleptonazation method that we employ in our simulations \citep{liebendoerfer:05}. In more realistic simulations with full neutrino transport, the differences in the bounce and ring-down GW signals from different progenitors should be far smaller \citep{mueller:09phd}.}

Apart from studying the accuracy metric and the confusion matrix, we also check the classification performance using the recall (also known as completeness) and precision (also known as correctness) of each progenitor mass class. The recall is defined as
\begin{equation}
    \mathrm{recall = \frac{TP}{TP + FN}},
    \label{eq:recall}
\end{equation}
while precision is defined as
\begin{equation}
    \mathrm{precision = \frac{TP}{TP + FP}},
    \label{eq:precision}
\end{equation}
where TP, FN and FP stand for the number of true positives, false negatives, and false positives for classification of target labels, respectively. Precision measures the quality of positive predictions made by the machine, i.e., the fraction of results which are relevant. Recall reflects the percentage of total true positive results correctly classified by the algorithm (hence also called as true positive rate or sensitivity). However, there is always a trade-off between recall and precision, and one can not maximize both these metrics at the same time \citep{recallprecision}.

\begin{figure}
    \includegraphics[width=\columnwidth]{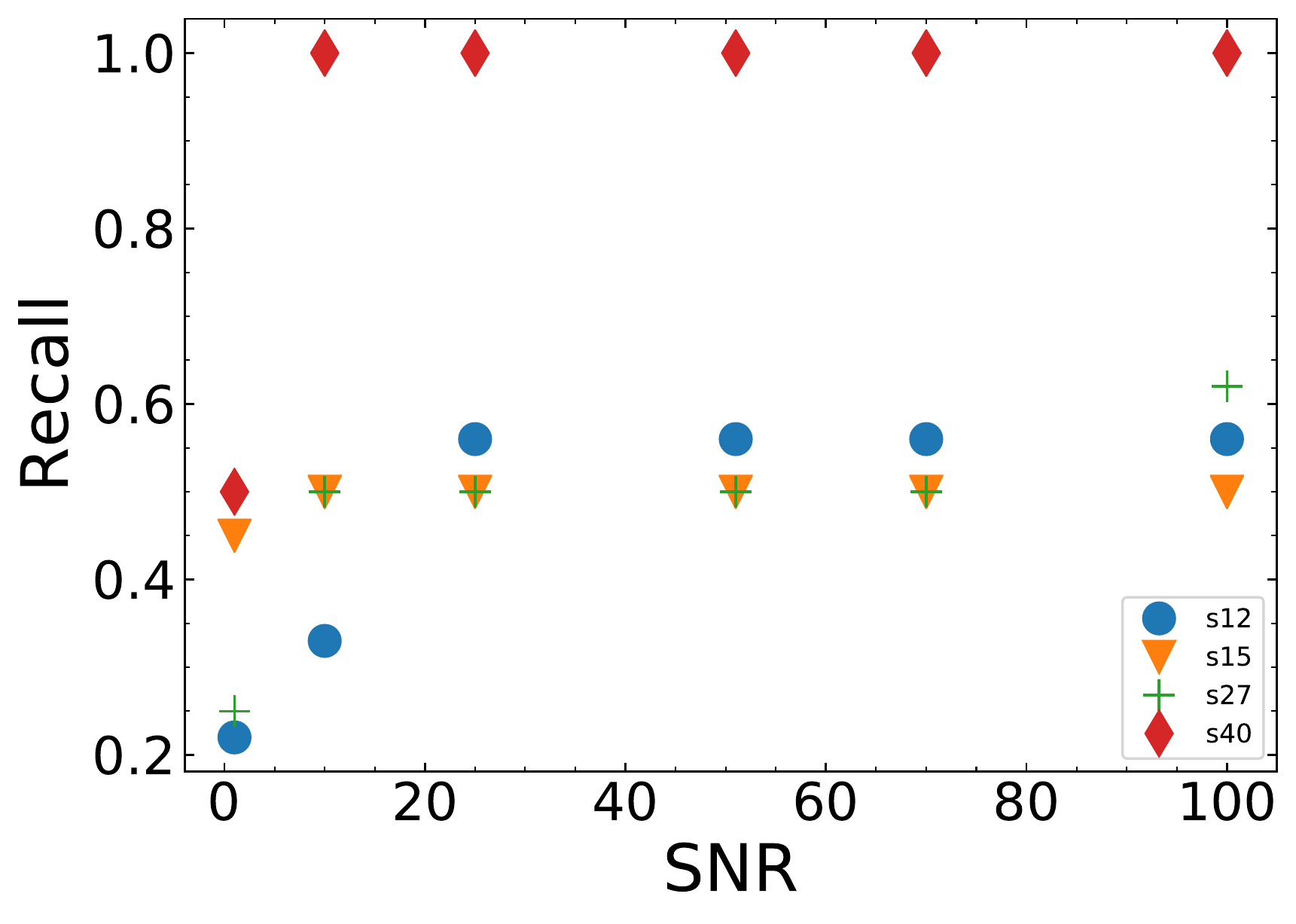}
    \caption{Recall as a function of the SNR for each progenitor mass. Recall gives insight into the number of positive cases that are misclassified by the model as negatives (i.e. number of false negatives). Higher recall value signifies higher amount of correctly classified positives. In this plot we can see that except for s40 model, recall values are below $0.7$. They only improve up to $\mathrm{SNR}\sim40$, except for s27 which shows a gain in recall at high SNR values. However, its corresponding  precision value falls in the same (high) SNR region (Fig.~\ref{fig:precision}) and thus nullifying any additional gain achieved in classification, by the improved recall performance in the high SNR range. }
    \label{fig:recall}
\end{figure}

\begin{figure}
    \includegraphics[width=\columnwidth]{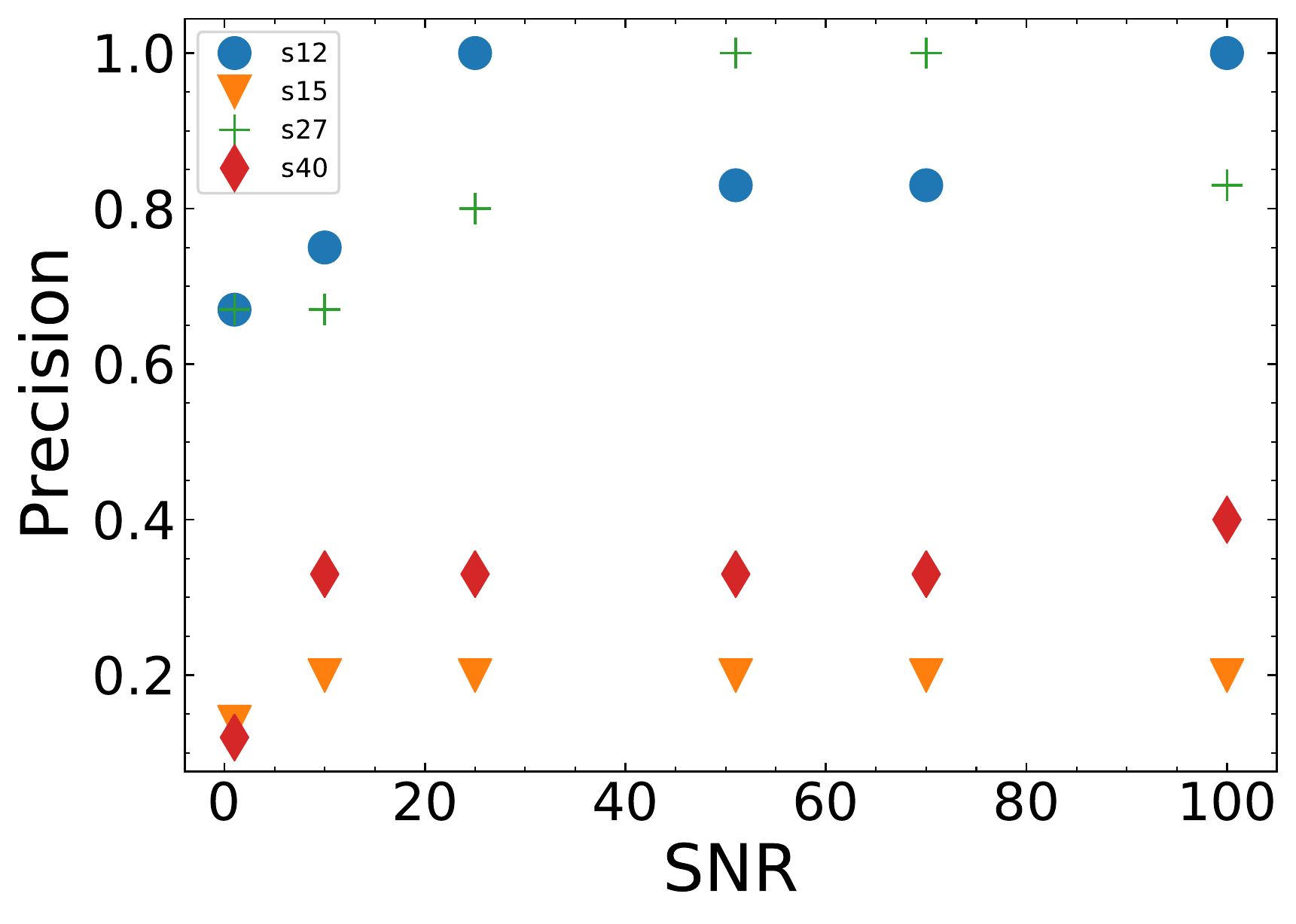}
    \caption{Precision as a function of the SNR for each progenitor mass. Precision gives insight into the quality of positive predictions made by the model. In this case it defines the number of labels the model correctly predicted divided by the total number of labels the model predicted. In this plot we see that beyond SNR$=40$, precision does not change significantly until it reaches maximum $\mathrm{SNR}=100$, where some improvement is observed apart from s27 model, which counteracts this behaviour by showing improvement in its recall values in the same SNR region (Fig.~\ref{fig:recall}). }
    \label{fig:precision}
\end{figure}

The recall and precision for our data are shown as a function of SNR in Fig.~\ref{fig:recall} and \ref{fig:precision}, respectively. We see from Fig.~\ref{fig:recall} that the s40 model achieves the highest recall. For $\mathrm{SNR}\gtrsim 15$, we see that recall value for s40 peaks off around $1$, implying zero or low false negatives for $\mathrm{SNR} \gtrsim 15$ (see Eq.~\ref{eq:recall}), making it the most successfully classified progenitor. This is also consistent with the values of the confusion matrix shown in Fig.~\ref{fig:Confusion}, which has an accuracy of $\sim 91\%$ for s40. The s15 model is seen to be the worst performing model. s12 and s27 show similar trends in their recall and precision curves. Except in model s27, where recall improves in high SNR range and correspondingly the precision falls, thus negating any additional gain in classification results, achieved by the improved recall at high SNR values. In conclusion, we find that the recall and precision plots in combination, show overall poor classification performance and also  beyond SNR $=40$ the improvement in performance levels off and minor gain is achieved.

\begin{figure*}
    \centering
    \makebox[1.\textwidth]{\includegraphics[width=1.0\textwidth]{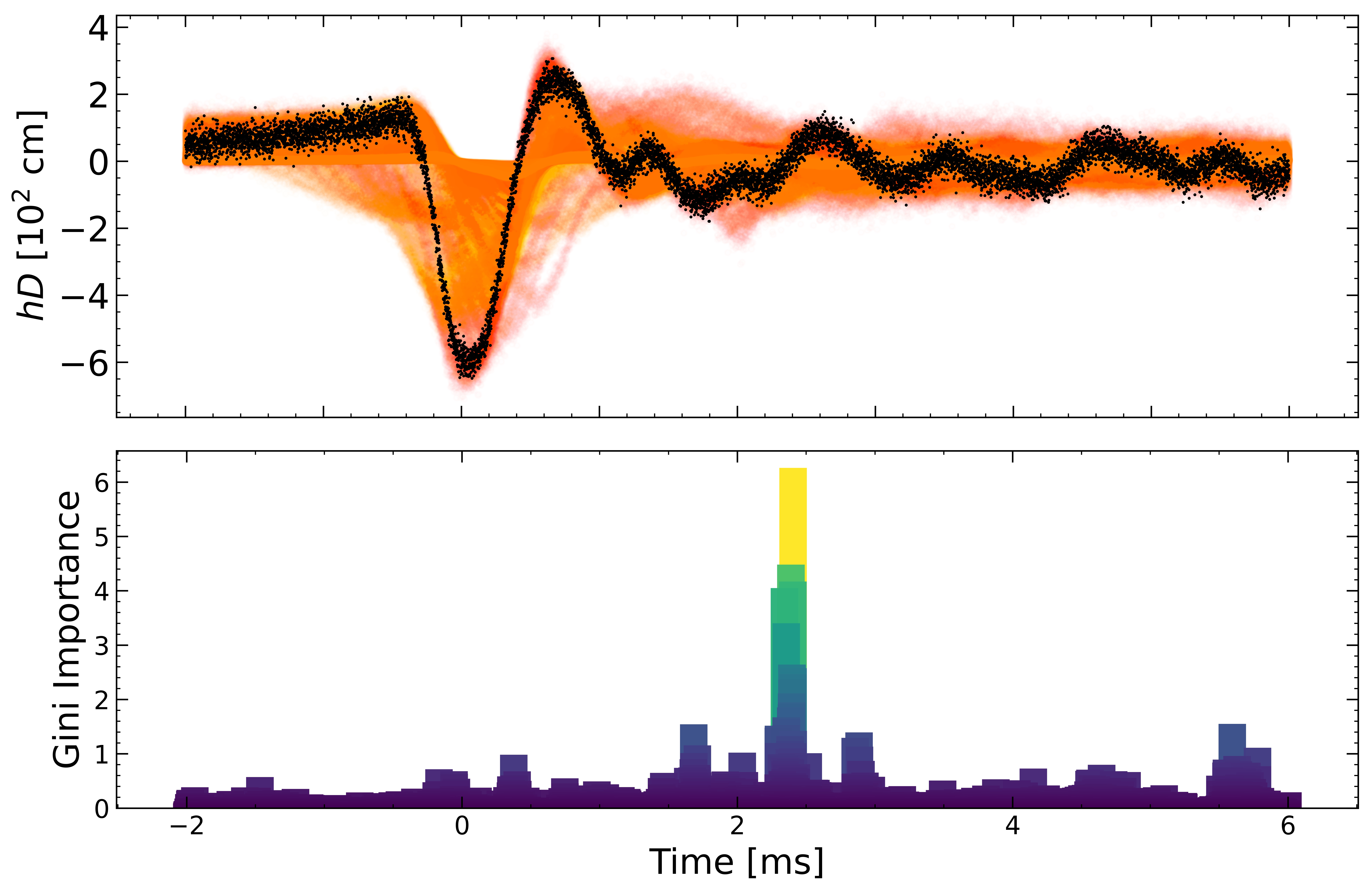}}
    \caption{GW strain for SNR $=15$ (top panel) and Gini importance score (bottom panel) plotted as a function of time. Highest importance score (scaled) is seen around $\sim 2.5$ ms after bounce. The Gini importance score indicates the features where maximum purity in Random Forest node splitting (selection process) occurred.}
    \label{fig:gini}
\end{figure*}

To understand the classification mechanism and which signal features the classifier labelled as most dominating, we use the Gini importance score or the Gini index \citep{Gini,Gini2}. The Gini index is commonly used as the splitting criterion in classification trees, the corresponding impurity importance is often called Gini importance or mean decrease in impurity. While it is debated that correlation between data features can hinder importance interpretation, it is still a robust metric for analyzing feature importance. Fig.~\ref{fig:gini} shows the Gini score mapped to the time axis of the input signals. It is seen that the GW signal features in the neighbourhood of $\sim\!2.5$ ms are regarded as most relevant by the classifier for splitting nodes across the whole tree. This signifies that in this region the maximum purity in node splitting occurs compared to the rest of the region in the waveforms. In this region, the GW signal is dominated by the ring-down oscillations of the PNS, suggesting that the difference in the initial mass influences the properties of PNS oscillations. However, Gini importance does have sensitivity to correlation of features which influences the quality of node splits along decision tree structures. Therefore higher score is obtained with uncorrelated signals or pure node splits. In our plot, the low Gini score except around $2.5$ ms implies that there must be correlation of features in GW signals, present in these regions. This might have reduced the purity during node splits and thus blurred identification between features.

To verify the robustness of these finding, we have explored sensitivity of the ML algorithm by performing our analysis with multiple different classifiers, including the XGBoost and Neural Network. As we show in Appendix~\ref{appendix-1}, the main finding of our paper remains the same: in all cases, the performance of the classifier is not satisfactory to recover target labels.

\section{Conclusion}

We have presented a detailed analysis of whether bounce and early ring-down GW signals from future observations of rotating CCSNe can be used for identifying iron core masses. We used GW data obtained from numerical simulations and injected white Gaussian noise to mimic the detector noise. We developed a ML classifier with the aim to identify the target classes of these input signals. We considered an idealized optimistic scenario for identifying the mass: we use rapidly rotating models that emit strong GWs, include only progenitor with only four different masses, which simplifies the selection process. In addition, to generate GW signals, we used numerical simulations that use a deleptonization scheme that artificially amplify differences in the collapse and post-bounce dynamics of progenitors with distinct masses. 

Despite this idealized favorable scenario, we were unable to identify the mass, purely based on bounce and early ring-down GW signal alone. More realistic treatment of noise modelling like non-Gaussian signals can only reduce the chances of any possible identification of the progenitor mass labels. The fact that even for SNR of 100, the classifier failed to give a high accuracy score, does point to the fact that this is an intrinsic phenomenon and independent of detector quality. Our results therefore show that the information about the iron core mass is not contained in the bounce and early ring-down GW signal. We have also performed similar analysis in Fourier space and classification performance was within $2\%$ of the accuracy score in time domain space. We thus conclude that the weak relation between GW waveforms and the iron core mass is an intrinsic property of the system and the classification via ML is not possible.

Additionally, we have explored sensitivity of the ML algorithm by performing our analysis with multiple different classifiers (e.g., XGBoost and Neural Network, as discussed in Appendix~\ref{appendix-1}). In all cases, our results showed that the performance of the classifier was not satisfactory to recover target labels. We do however note that, as with any ML analysis, numerous other hyperparameter combinations are possible, as do multiple other algorithms. We anticipate to further perform our analysis in future with more complex ML model and representative data. However, we do not expect significant deviation from the results presented in this paper. 

{ While our work show that the bounce and early ring-down GW signal do not contain information about the iron core mass, this does not mean that the iron core or progenitor mass cannot be measured from GWs at all. The explosion dynamics depends sensitively on the progenitors structure \cite[e.g.,][]{oconnor11, ugliano:12, mueller:16c} and it is imprinted in the GW signal emitted in the corresponding phase \cite[e.g.,][]{mueller:13, radice:19gw}. This signal can be used to constrain the progenitor mass \citep{pajkos21}. Additionally, incorporation of the neutrino signal will further enhance the measurements \citep[e.g.,][]{oconnor:13, Yokozawa15, kuroda:17, Nagakura22}.} 

\section*{Acknowledgements}

We thank Sherwood Richers { and David Vartanyan} for carefully reading the manuscript and for helpful comments. This research has been funded by the Science Committee of the Ministry of Education and Science of the Republic of Kazakhstan (Grant No. AP13067834 and AP08856149) and the Nazarbayev University Faculty Development Competitive Research Grant Program No 11022021FD2912. 
High-performance workstations of ECL/NU have been used to perform all simulations, data analysis, and ML calculations (\href{http://ecl.nu.edu.kz/computational-facilities/}{http://ecl.nu.edu.kz/computational-facilities/}).

\section*{Data Availability}

The data used in this work is available from authors upon request. The gravitational waveforms are publicly available at \href{https://zenodo.org/record/7090935}{https://zenodo.org/record/7090935}.


\bibliographystyle{mnras}
\bibliography{references} 



\appendix

\section{Principal Component Analysis}
\label{appendix:PCA}

Principal component analysis (PCA) is a well-known method for dimension reduction generally applied before a full scale ML analysis \citep{PCA}. It was used in several works in the context of CCSN GWs \citep[e.g.,][]{Rover:09, Logue12, Engels14, Edwards14, Afle21, Suvorova19}. In our work, we have the advantage of having a relatively small data size by ML standards and the availability of high computational resources without any time constraint. This means we can run a full scale ML analysis without the requirement to apply a dimension reduction technique, before classification stage. We still performed a PCA analysis of our data for further insight. 

\begin{figure}
    \includegraphics[width=\columnwidth]{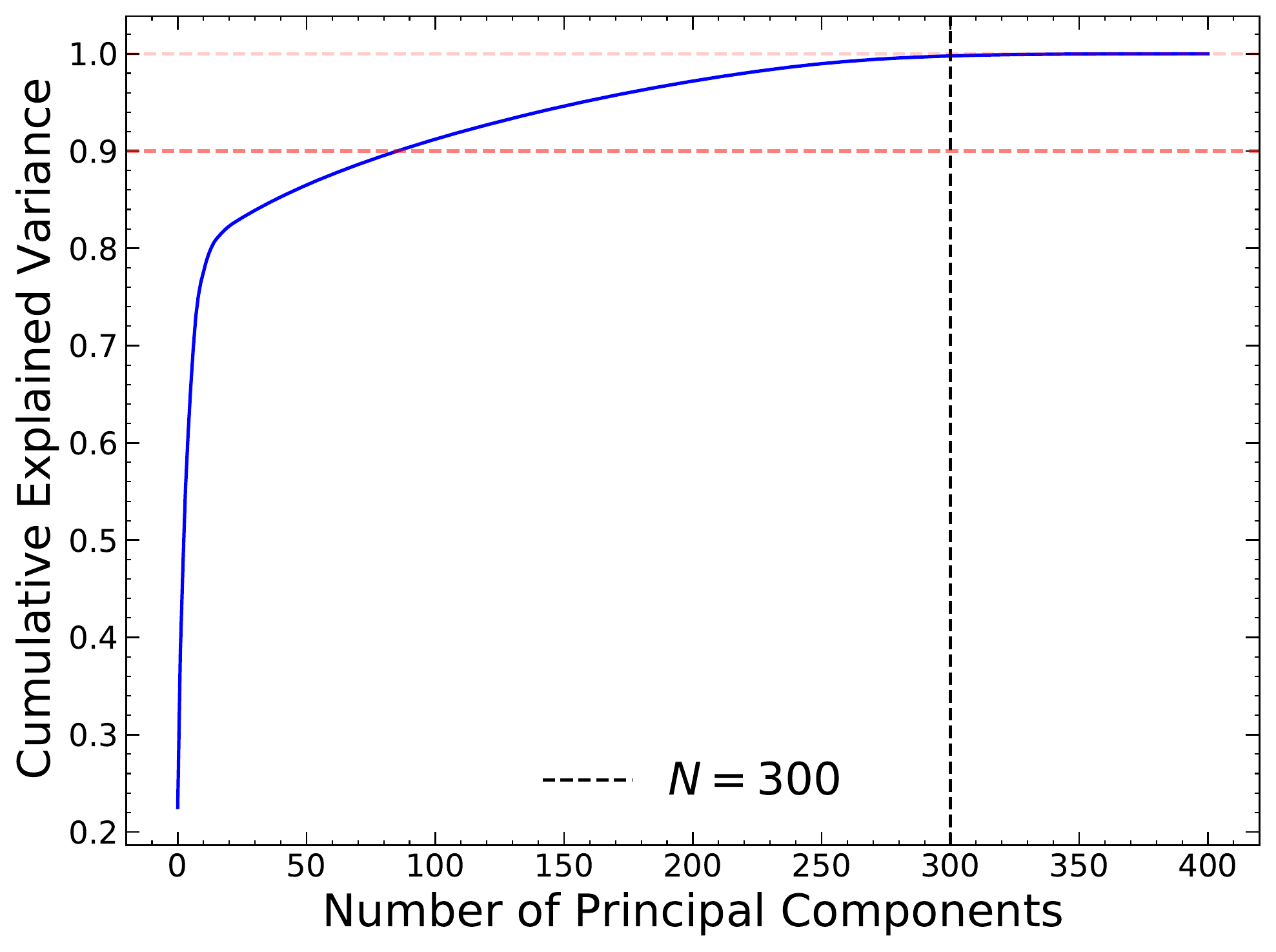}
    \caption{Amount of variance captured as a function of the number of principal components. The red dashed horizontal lines represent the $90$ and $100\%$ cumulative variance mark while the black dashed vertical line corresponds to $N=300$ which is the mark where cumulative variance reaches $\sim 1$. }
    \label{fig:pca}
\end{figure}

We illustrate our findings in Fig.~\ref{fig:pca}, which shows the eigenvalues ($y$-axis) as a function of the number of components along the $x$-axis. It helps us to estimate the reduced dimension of the data set with minimal loss in accuracy. The principal components are constructed from the linear combination of eigenvectors obtained from the covariance matrix of the data. The principal components are uncorrelated. The maximum variance (information) is carried by the first component, which decreases along the successive components. In other words, the first few principal components help us capturing most of the information of the data. We show that about $75\%$ of the data, which corresponds to the number of principal components of $\simeq 300$, is adequate to capture the entire variance, as seen along the $y$-axis of the plot with cumulative variance $\simeq\! 1$. Furthermore, only $25\%$ of the data, which is represented by $\simeq\!100$ by principal components, is sufficient to capture $90\%$ of the total variance in the input data. This can speed up the calculations by a few orders magnitude compared to applying RF on the full size data set. For future analysis with potentially more complex algorithms (e.g., long short term memory \citep{lstm}) and for quicker reproduction of results in a bigger survey pipeline, applying a PCA can prove to be necessary.   

\section{Alternative Algorithms}
\label{appendix-1}

The analysis presented in this paper is based on a RF classifier, which performed best compared to the other choices that we considered. In this section, we will summarize the two other algorithms that we tried: the XGBoost and Neural Network based classifiers.

\subsection{XGBoost Classifier}\label{sec:xgboost}

The Extreme gradient Boosting classifier, in short XGBoost classifier, is a tree optimisation technique based on gradient boosting \citep{xgb, gradientree}. If the output of an ensemble of $K$ trees is given as 
\begin{equation}
    y_i = \sum_{k=i}^{K}f_k(\chi_i),
\end{equation}
such that each $f(\chi)=w_q(\chi_i)$, where $x$ is the input vector and $w_q$ is the score of the corresponding leaf $q$, then the objective function $J$ that the XGBoost algorithm tries to minimise at each step $t$ is
\begin{equation}
    j(t)=\sum_{i=1}^{n}\left(y_i,\hat{y}^{t-1}+f_t(x_i) \right)+\sum_{i=1}^{t}\Omega(f_i).
\end{equation}
The first term $L$ contains the train loss function of mean squared error between the true class $y$ and the predicted class $\hat{y}$ for the $n$ samples. The last (second) term on the right is for regularization term controlling the model complexity and preventing an overfit. The second term $\Omega(f)$ also defines the model complexity and can be expressed as
\begin{equation}
    \Omega(f)=\gamma T + \frac{1}{2\lambda\sum_{j=i}^T \left(w_j^2\right)},
\end{equation}
where $T$ is the number of leaves, $\gamma$ is the pseudo-regularization hyperparameter, which is dependent on each dataset. Parameter $\lambda$ is the L2 norm for leaf weight.

With the XGBoost classifier, we follow the exact same pipeline as with the RF classifier except tuning for the XGBoost specific hyperparameters in the grid search stage. The hyperparamters used in this algorithm is listed in Table~\ref{tab:xgb}.

\begin{table}
        \setlength\extrarowheight{2.09pt}
        \begin{center}
            \begin{tabular}{|c| }
                \hline
                Hyperparameters   \\ [.3em]
                \hline
                No. of estimators   \\ 
                Learning rate   \\ 
                Max. depth   \\
                Booster   \\
                reg$(\alpha)$  \\
                reg$(\lambda)$   \\
                Base\_score   \\
                \hline
            \end{tabular}
        \end{center}
        \caption{Hyperparameters that are tuned in the grid search step in the XGBoost classification analysis.}
        \label{tab:xgb}
    \end{table}
    
\begin{figure}
    \includegraphics[width=\columnwidth]{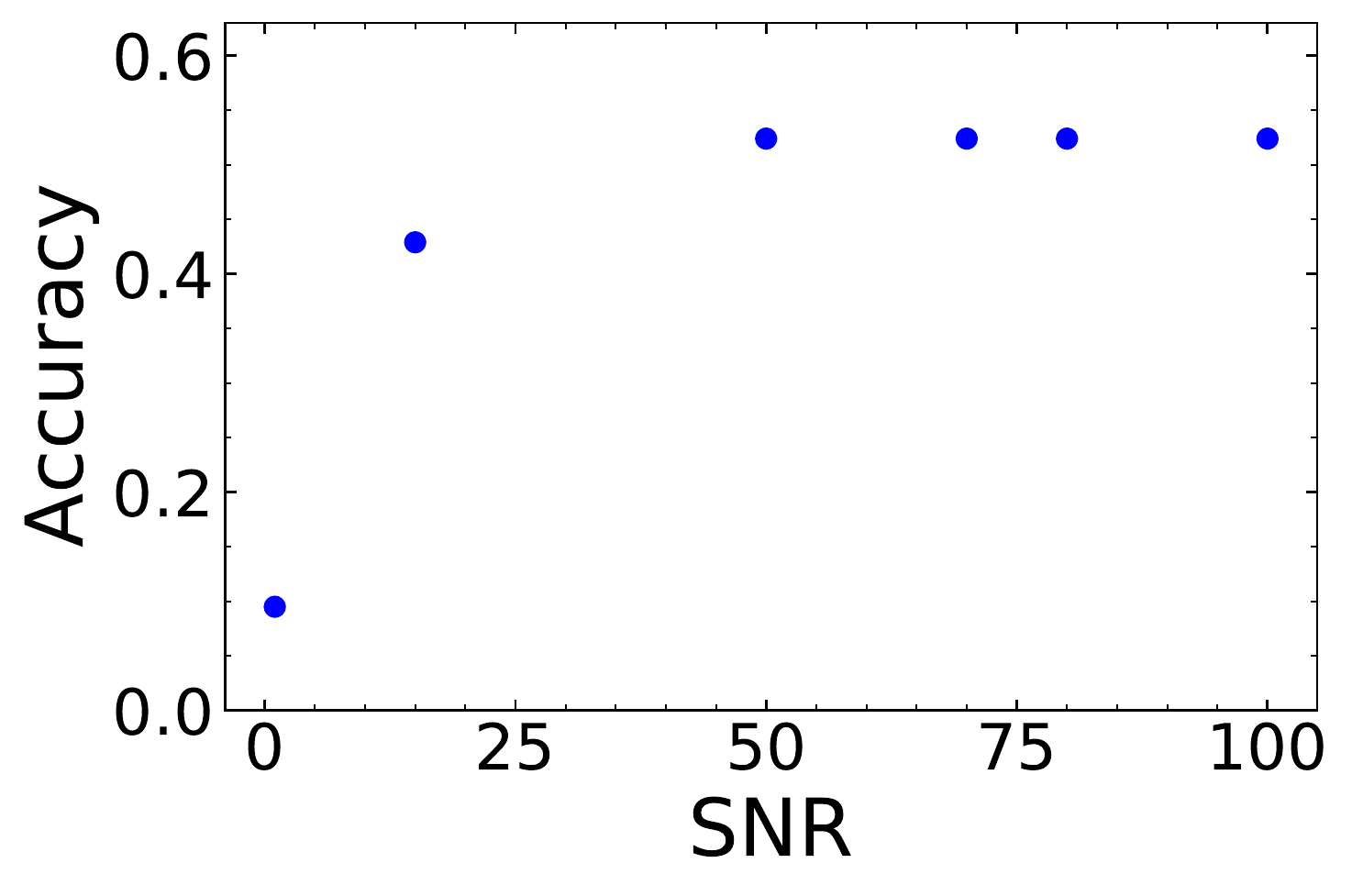}
    \caption{Accuracy as a function of SNR with XGBoost classifier, similar to Fig.~ \ref{fig:accuracy}. It is seen that beyond SNR$=50$, the accuracy score does not improve.}
    \label{fig:Accuracyxbg}
\end{figure}

Fig.~\ref{fig:Accuracyxbg} illustrates the accuracy values as a function of the SNR, similar to Fig.~\ref{fig:accuracy}. We can see that for $\mathrm{SNR}=100$, i.e. the highest value used in this analysis, the accuracy is $\sim50\%$. This value is significantly smaller compared to RF classifier's accuracy score of $\sim70\%$, shown in Fig.~\ref{fig:accuracy}. Since both XGBoost and Random Forest classifiers are some of the most robust algorithms, this finding reinforces our finding that classifying the core mass seems impossible from the bounce and early ring-down GW signal alone. 

\subsection{Deep Learning: Neural Network Classifier}

\begin{figure}
\centering
\includegraphics[width=\columnwidth]{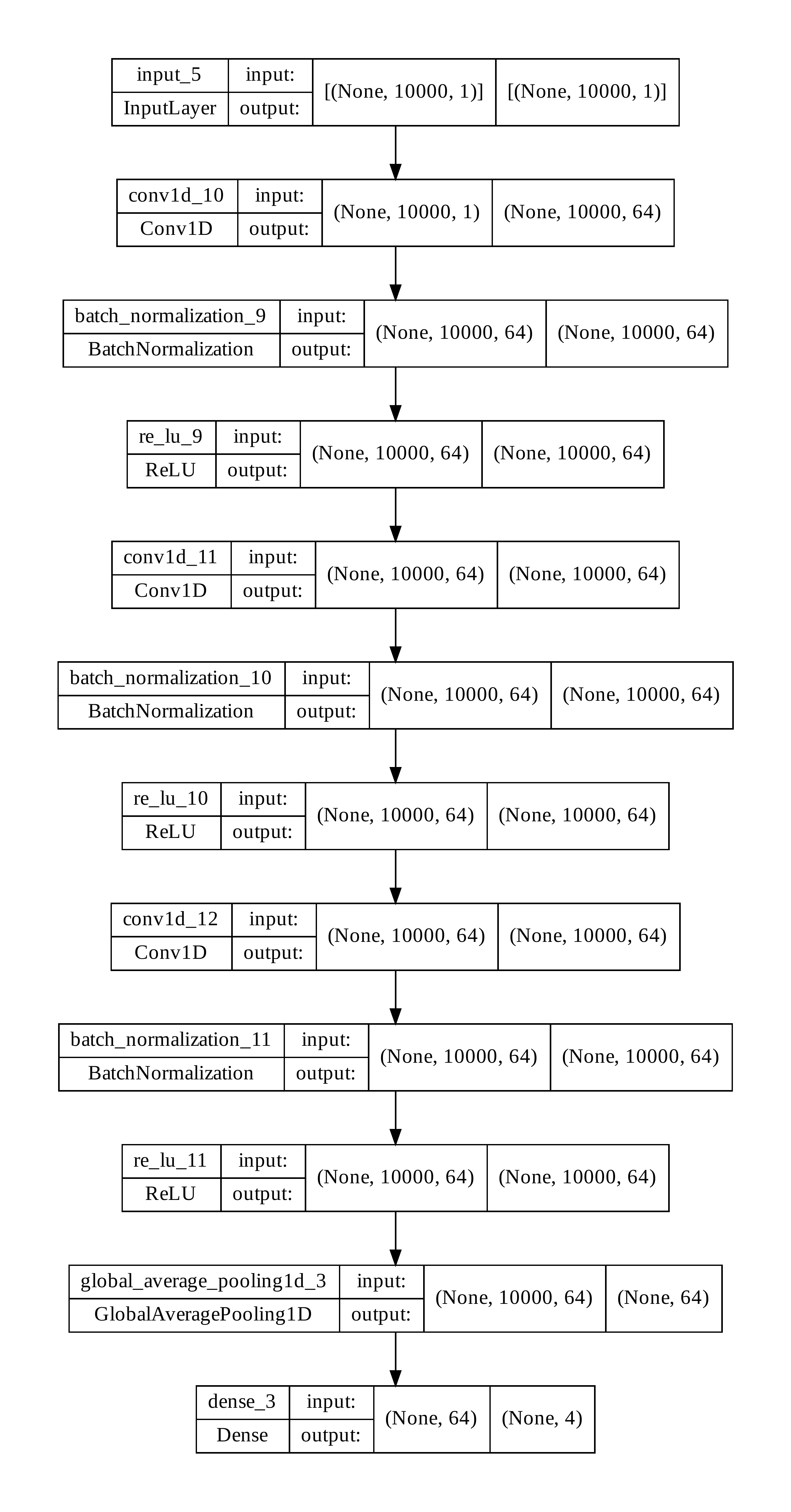}
\caption{Architecture of the fully connected neural network used in this analysis based on \citet{NN}.}
\label{fig:NN}
\end{figure}

We run a separate convolutional neural network (CNN) based time series classifier algorithm on our data, based on the architecture presented in \citet{NN}. They provided an end-to-end fully connected time series classification architecture without any need for feature engineering and data processing. This architecture was shown to be successful with such time series classification, outperforming all other neural network based architectures. This motivates a classification analysis of our data with this pipeline. A flowchart cartoon of the architecture is illustrated in Fig.~\ref{fig:NN}. 

\begin{figure}
\centering
\includegraphics[width=\columnwidth]{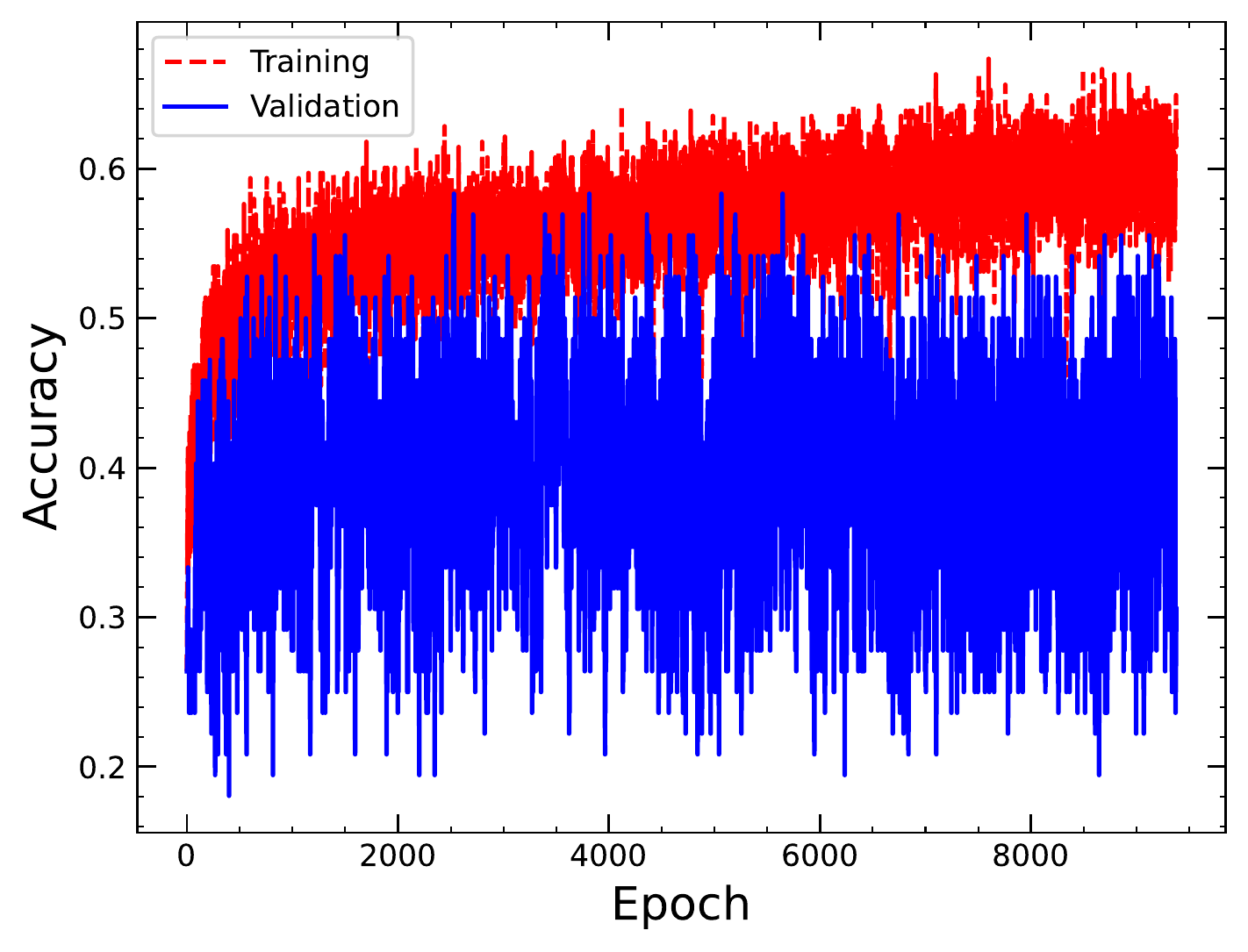}
\caption{Accuracy score for the neural network for $\mathrm{SNR}=100$. The highly noisy pattern of the validation curve suggests that the model heavily overfits.}
\label{fig:NNAccuracy}
\end{figure}

We assess the model performance with sparse categorical crossentropy metric. As seen from the training and validation curve  in Fig.~\ref{fig:NNAccuracy}, the model heavily over-fits with the given architecture. The overfit can be seen from the noisy distribution of points in the curves. For the best case scenario of $\mathrm{SNR}=100$, we achieve a classification accuracy of $\sim 55\%$ (Fig.~\ref{fig:NNAccuracy}). From this, we conclude that the neural network pipeline performance is consistent with our RF classification results, but the network still needs significant improvement in its design. In our case, we expect a simpler architecture could reduce the over-fit problem.

\bsp	
\label{lastpage}
\end{document}

%% file: formula.tex
\newcommand{\window}{$[-2,\ +6]$ ms}
\newcommand{\total}{$402$}

\newcommand{\mrm}{\mathrm}

\newcommand{\be}{\begin{equation}}
\newcommand{\ee}{\end{equation}}
\newcommand{\bea}{\begin{eqnarray}}
\newcommand{\eea}{\end{eqnarray}}

\definecolor{mediumpurple}{rgb}{0.58, 0.44, 0.86}